\documentclass[]{aa}  
\newcommand{\cd}{d$^{-1}$}
\newcommand{\kms}{km\,s$^{-1}$}
\newcommand{\ms}{m\,s$^{-1}$}

\newcommand{\vsini}{$v\sin{i}$}

\usepackage{graphicx}
\usepackage{txfonts}
\usepackage{natbib}
\begin{document}
   \title{First evidence of pulsations in Vega?
      \thanks{Based on observations obtained at the Bernard Lyot Telescope (TBL, Pic du Midi, France) of the Midi-Pyr\'en\'ees Observatory, which is operated by the Institut National des Sciences de l'Univers (INSU) of the Centre National de la Recherche Scientifique of France (CNRS), and at the Canada-France-Hawaii Telescope (CFHT), which is operated by the National Research Council of Canada, INSU/CNRS and the University of Hawaii.}}
   \subtitle{Results of today's most extensive spectroscopic search.}

   \author{%
          T. B\"ohm\inst{\ref{inst:irap1},\ref{inst:irap2}} \and
          Fran\c{c}ois Ligni\`eres\inst{\ref{inst:irap1},\ref{inst:irap2}} \and 
          Gregg Wade\inst{\ref{inst:royalmil}}\and
          Pascal Petit \inst{\ref{inst:irap1},\ref{inst:irap2}}\and
          Michel Auri\`ere\inst{\ref{inst:irap1},\ref{inst:irap2}}\and
          Wolfgang Zima\inst{\ref{inst:sterre}}\and         
          Aur\'elie Fumel\inst{\ref{inst:irap1},\ref{inst:irap2}} \and
          Evelyne Alecian\inst{\ref{inst:lesia}}
          }

   \offprints{T. B\"ohm}

   \institute{ 
              Universit\'e de Toulouse; UPS-OMP; IRAP; Toulouse, France\label{inst:irap1}\\
              \email{boehm@obs-mip.fr}
       \and       
              CNRS; IRAP; 14, avenue Edouard Belin, F-31400 Toulouse, France\label{inst:irap2}
       \and
          Instituut voor Sterrenkunde, Celestijnenlaan 200D BUS 2401\label{inst:sterre}
       \and
           Dept. of Physics, Royal Military College of Canada, PO Box 17000, Stn Forces, Kingston, Canada K7KK 7B4\label{inst:royalmil}
        \and
          Observatoire de Paris-Meudon, LESIA, 92190 Meudon, France\label{inst:lesia}
     }

   \date{Received September 27, 2011; accepted xx}

 
  \abstract
   {The impact of rapid rotation on stellar evolution theory remains poorly understood as of today. Vega is a special object in this context as spectroscopic and interferometric studies have shown that it is a rapid rotator seen nearly pole one, a rare orientation particularly interesting for seismic studies. In this paper we present a first systematic search for pulsations in Vega .}
   {The goal of the present work is to detect for the first time pulsations in a rapidly rotating star seen nearly pole-on.}
   {Vega was monitored in quasi-continuous high-resolution echelle spectroscopy. A total of 4478 spectra were obtained within 3 individual runs in 2008, 2009 and 2010. The resolution was higher than R =  65000. This data set should represent the most extensive high S/N, high resolution quasi-continuous survey obtained on Vega as of today.
Equivalent photospheric absorption profiles were calculated for the stellar spectrum, but also the telluric lines present in the spectrum, the latter were used as a radial velocity reference. Residual velocities were analysed and periodic low amplitude variations, potentially indicative of stellar pulsations, detected. In a subsequent step, LSD profile variations were searched for in a bidimensional analysis night by night. Finally, a Lomb-Scargle periodogram of each velocity bin of the profile was performed for all three observing runs.}
   {Based on high resolution echelle spectroscopy, we have obtained indications of periodic variations of very small amplitudes within the residual radial velocity curves of Vega. All three data sets revealed the presence of residual periodic variations: 5.32 and 9.19\,\cd\, (A $\approx$ 6\ms) in 2008, 12.71 and 13.25\,\cd\, (A $\approx$ 8\ms) in  2009 and 5.42 and 10.82\,\cd\, (A $\approx$ 3-4\ms) in 2010. However, it is too early to conclude that the variations are due to stellar pulsations, and a confirmation of the detection with a highly stable spectrograph is a necessary next step.}
   {If pulsations are confirmed, their very small amplitudes show that the star would belong to a category of very "quiet" pulsators.}

 
   \maketitle
%

\section{Introduction}
Rapid rotation is a characteristic feature of massive and intermediate mass stars
that has a significant and yet poorly understood impact on their structure and evolution.
It is a current challenge of stellar physics to obtain observational constraints
that will improve the modeling of the physical processes induced
by rotation. Seismology should play a particular role in this context as it
is the only way to directly probe stellar interiors.

At present, however, the oscillation spectra of rapidly rotating stars can not be
unambiguously interpreted. One difficulty resides in the modeling
of the effects of rapid rotation on the oscillation modes.
There has been some recent progress in this field notably in the description
of the organization of p-modes spectra  \citep{lingor}.
Even if we understand the basic properties of the frequency spectrum,
the mode identification process remains a difficult task.
In particular, in the p-mode frequency range, the large rotational splittings between
modes of different azimuthal number mix the frequencies of different multiplets.
One possible solution to this problem would be to observe a rapidly rotating pulsating star seen pole-on. Cancellation effect
in disk-integrated light would then select axisymmetric modes, non-axisymmetric modes being cancelled out.
Such a selection effect would considerably simplify the observed spectrum and thus the mode identification process.

Vega is known to be a rapidly rotating star and reveals a nearly pole on inclination. This has been shown
by analyzing the spectroscopic and interferometric signatures of the surface gravity darkening 
 \citep{gulli,peterson}.
The spectroscopic and interferometric approaches provide a similar inclination angle  i = 5-7 degrees, 
although they differ in the reported equatorial velocities, v$_{\rm eq} = 245 \pm 15$\kms\, and v$_{\rm eq} \sim$  175 \kms\, from the spectroscopic analysis of \citet{gulli} and \citet{Ta08}, respectively, 
and a significantly larger value, v$_{\rm eq} \sim$  275 \kms, from interferometry \citep{peterson,aufden}.
As a seismic target, Vega also presents the important advantage
that its fundamental parameters are very accurately determined, particularly when compared
with other rapidly rotating stars.

As reviewed by \citet{Gr07}, Vega was used since more than 150 years as a photometric and spectrophotometric standard star, 
the historic reason being most likely its brightness and near-zenith position for european and north american observatories.  
This particular role of Vega can be considered surprising, since several authors report photometric variability of at least $1-2 \%$,  but with no evidence of periodic variations.
A previous study of spectroscopic variability has been conducted by \citet{goraya} who described a variability of H$\alpha$ althought
it was not confirmed in subsequent studies \citep{charlton}.

The location of Vega in the HR diagram is roughly that of main-sequence A0 stars, but bluewards of the $\delta$ Scuti instability strip, therefore stellar oscillations are in principle not expected. However, \citet{fernie} concluded that historic radial velocity curves and photometric variability tend to indicate episodical 
$\delta$ Scuti type pulsation, with a frequency close to the fundamental radial order of F$_{\rm fund} \sim $10\,\cd\, for Vega's parameters as known at that time. The most recent determination of fundamental stellar parameters by \citet{Ta08} led him to conclude that this rapid rotator has parameters strongly varying from pole to equator, more precisely an effective temperature from $T_{\rm eff}$ = 9867 to 8931\,K, a surface gravity from $\log g$ = 3.308 to 3.313 (where $g$ is expressed in cm\,s$^{-2}$) and a radius from $R$ = 2.520 to 2.763 R$_{\odot}$, respectively. 

Until know, no rapidly rotating, pole-on and observably pulsating star was known. The goal of our work was therefore to search for low amplitude stellar pulsations in an extensive high-resolution quasi-continuous spectroscopic survey of Vega.

Sect. \ref{obs} describes the observations and data reduction, Sect. \ref{searchpulsations} presents the analysis of radial velocity variations in the spectroscopic time series of Vega, and in Sect. \ref{disc} results are discussed and
a conclusion is drawn.


\section{Observations and data reduction}
\label{obs}

\begin{table*}
\caption[]{Log of the spectroscopic (\& polarimetric) observations of Vega. (1) Date  of the observation (UT) (\citet{vegamag2} indicates the date in local time, leading to a -1 day shift for the Sept. 2009 CFHT data). (2)  and (3) Heliocentric Julian date (mean observation, 2,450,000+) of the first and the last stellar spectrum of the night, respectively; (4) Number of high resolution RS Cha spectra; (5) exposure time (sec);(6) total hours covered on the sky; (7) nightly average and standard deviation of signal to noise per resolved element at 520\,nm.}
\begin{tabular}{ccccccl}
\hline\hline
 Date	& JD$_{\rm first}$	& JD$_{\rm last}$	& N$_{\rm spec}$	& t$_{\rm exp}$ (sec)& t$_{\rm cov}$ (hrs) & S/N \\
(1)		&  (2)  			&  (3)        			&   (4)         		&        (5)                       &     (6)                        &   (7)  \\
\hline
Jul. 26  2008 &  4674.3489	&	4674.6072	& 389	                   &	6			&   6.2     			& 812$\pm$50\\ 
Jul. 27  2008 &  4675.3748       &       4675.6551        & 390                           &	6			&   6.7     			& 822$\pm$61\\
Jul. 29  2008 &  4677.3480       &       4677.6415        & 434                           &	6			&   7.0    			& 823$\pm$82\\ \hline
Sep. 9   2009  &   5083.7400	&      5083.9066         & 402                           &	4			&   4.0   			& 865$\pm$85\\
Sep. 10 2009  &   5084.7111       &       5084.8787        & 331                           &	4			&   4.0     			& 887$\pm$69\\
Sep. 11 2009  &    5085.7122    &      5085.9498         & 560                           &	4			&   5.7      			& 736$\pm$155\\ \hline
Jul. 15  2010 & 5393.3643         &      5393.5735	& 355	                   &      13                       &   5.0                          & 1117$\pm$37\\
Jul. 16  2010 & 5394.3635         &      5394.6576	& 496	                   &      13                       &   7.1                          & 1051$\pm$158\\
Jul. 17  2010 & 5395.3650         &      5395.6510	& 453	                   &      13                       &   6.9                          & 1197$\pm$59\\
Jul. 18  2010 & 5396.3580         &      5396.4662	& 185	                   &      13                       &   2.6                          & 1222$\pm$16\\
Jul. 19  2010 & 5397.3561         &      5397.6569	& 483	                   &      13                       &   7.2                          & 1170$\pm$70\\ \hline
\end{tabular}
\label{table:log}
\end{table*}


The analysis presented in this paper is based on three datasets obtained in 2008, 2009 and 2010. Vega was monitored 
during 3 nights in 2008 (26$^{\rm th}$, 27$^{\rm th}$ and 29$^{\rm th}$ of July) with the high-resolution echelle spectropolarimeter NARVAL at the 2m Bernard Lyot Telescope (TBL, Pic du Midi, France). The first two nights observations were acquired at R = 65000 in polarimetric mode, and the last night at R = 75000 in spectroscopic mode. We obtained 1213 spectra covering 19.9 hrs. In 2009 we obtained 1293 spectra covering 13.7 hrs on 3 nights (9$^{\rm th}$ - 11$^{\rm th}$ of September 2009) at ESPaDOnS/CFHT (R = 68000) in polarimetric mode. Finally, in 2010, we got a huge data set of 1972 spectra covering 28.8 hrs on 5 nights (15$^{\rm th}$ - 19$^{\rm th}$ of July 2010), obtained again in the polarimetric mode at NARVAL/TBL. Table \ref{table:log} summarizes the log of the observations during these 3 runs. 

The general observing strategy was to obtain as many maximum S/N observations of the target during the night as possible. The goal was to obtain a versatile data set, enabling us to search for the presence of a magnetic field as well as stellar pulsations.  Results of the spectropolarimetric study of the first observing run are presented in \citet{vegamag}, where the detection of a weak magnetic field in Vega has been announced, revealing the existence of an as-yet unexplored class of weakly magnetic stars in the intermediate mass domain. The joint analysis of the two first runs (2008 and 2009) confirmed this first detection and indicated a rotational modulation of the polarization signature with a period of 0.732 $\pm$ 0.008\,d, a value which is very close to the rotation period announced by \citet{Ta08} ($\simeq$ 0.733\,d). 
These results are published in \citet{vegamag2}. A recent analysis of the 2010 data set yielded a slightly different rotation period of 0.678$^{+0.036}_{-0.029}$ d \citep{alina2011}.

The global data set containing 4478 spectra obtained at the highest possible signal to noise should represent, as of today, the most extensive 
high resolution quasi-continuous spectroscopic monitoring of Vega. It should be noted that for such a bright star the exposure times are extremely short (4 to 13 seconds on the star) in order to avoid the saturation of the CCD, while the readout times of the CCD dominate the duty-cycle, the average time interval between two exposures being close to 0.8 min. Subexposures obtained in polarimetric mode provide as a direct result from the data reduction pipeline, in addition to the polarization information, a perfectly calibrated intensity spectrum. 

The wavelength covered extends from 369 to 1048\,nm for both spectrographs. Standard calibration spectra were obtained at the beginning and at the end of the night; most importantly the absolute wavelength calibration relies on a Th/Ar spectrum at the beginning of the night. Most of the data reduction was carried out following standard reduction procedures using a dedicated
spectroscopic data reduction package, based on the algorithm described in  \cite{donati}. This package also makes use of the ``optimal extraction algorithm" \citep{horne}. A standard heliocentric velocity correction was performed. The intrinsic wavelength calibration accuracy achieved with the 2D-polynomial fit procedure is better than 150\,ms$^{-1}$ for both data sets. In order to be able to calibrate potential spectral shifts during the night a cross-correlation of telluric lines present in the spectra was performed by the standard reduction package. 

After all shifts have been applied to the wavelength grid, the residual telluric line velocity should reflect precisely the opposite of the heliocentric velocity correction. In order to verify this, we computed  least square deconvolved (LSD) profiles \citep{donati} of the more than 100 telluric water vapor lines contained in each stellar spectrum, and determined their centroid by fitting a gaussian profile. We discovered the presence of residual radial velocity errors of the telluric lines. In order to optimize the radial velocity precision we corrected then for the remaining velocity shifts. The residual spectral stability is around 10-15\,ms$^{-1}$ (depending on the night and on the site), taking into account all the multiplex information of the lines present in the spectra. 

At this point it should be noted that a stability analysis  of O$_2$ telluric lines has been carried out by \citet{caccin}. They report that the line positions change more or less erratically with time, their displacement depending upon the horizontal wind structure above the observing site. They note however, that unless the observations are made near the horizon, the component of the wind velocity along the line of sight is very seldom significant, the displacement reaching up to 10\,ms$^{-1}$ only in very severe conditions (large air mass, strong winds). In low wind conditions and for only one to three airmasses their time instability should be probably of the order of 5\,ms$^{-1}$. This is the case for our observations of Vega, since they have been obtained in airmasses between 1 and 2, and only during the last night of the CFHT run airmasses up to 2.6 were reached. In a recent work, \citet{figueira} evaluated the stability of atmospheric lines with HARPS (on the 3.6\,m telescope at La Silla, ESO), and concluded in a similar way to \citet{caccin} that a precision of 5\,ms$^{-1}$ can routinely be obtained in restricted airmass conditions. What we retain from this latter work is the fact that increasing airmass induces asymmetry of telluric lines, this asymmetry is therefore expected to have an impact on the measured RV. This behaviour-depending on the airmass of the observed object repeats therefore every night for the same target (within a 3 or 5 night observing run). Any periodic behaviour with different periodicities than 1d cannot be attributed to this intrinsic telluric line variability. Moreover, even if  \citet{caccin} invoke high altitude atmospheric movements or pressure shifts as responsible for inducing asymmetry in telluric lines, we could not find any report of  sinusoidal radial velocity variation of telluric lines during an observing night.

The next step of the data reduction was to calculate for all stellar spectra photospheric LSD profiles,
using a mask corresponding to the spectral type A0 (see also \citet{vegamag2}). We have to notice that, as \citet{gulli} and \citet{Ta08} have shown, the spectrum of this rapidly rotating star is very inhomogeneous, since the weak spectral lines form exclusively in the almost 1000\,K colder gravity darkened equatorial zone, leading to a flat-bottomed trapezoidal profile shape. However, the more numerous strong lines, and consequently the mean LSD profile, come from intermediate longitude to polar regions; therefore an A0 mask represents a good choice for Vega. 

The equivalent photospheric profiles contains multiplex information from more than 1100 individual photospheric lines. Typical S/N values of the resulting LSD intensity (Stokes I) profiles are consistently around 1000 per resolved element, following the behaviour pointed out by \citet{donati}, and indicating that the convolution model underlying the LSD cannot be trusted above this level of accuracy (this limitation is not valid for the other Stokes profiles).  Despite its almost pole-on geometry of i = 7$^{\circ}$ \citep{Ta08}, the LSD profiles of Vega  (\vsini\, = 22 \,kms$^{-1}$) needed to be fitted by  rotationally broadened profiles in order to perform an accurate radial velocity determination. The finaly achieved precision in radial velocity is again 10-15\,ms$^{-1}$ for each stellar LSD profile of this data set. 
The only way to take into account a potential instrumental drift during the night is to work on the difference between stellar and telluric lines, knowing the limitations of the telluric lines as an absolute radial velocity reference.

\section{Search for periodic variations in Vega}
\label{searchpulsations}

\subsection{Radial velocity variations in Vega}

Due to the one year time gap between the three runs we decided to analyse each run separately. Plotting each radial velocity time series separately, we observed the presence of trends within some nights of less than 100 \ms/night for the TBL 2008 run, and less than 60 \ms/night for the CFHT 2009 and TBL 2010 run. These slow variations did not appear to be sinusoidal, and we decided to take out these largescale trends by normalising the radial velocity curve of each night by a 2nd degree polynomial. In order to check for consistency, we analysed all three couples of normalized and raw radial velocity curves and concluded on the coherence of this approach by detecting the same frequencies in the range above 5\,\cd\,. 

\begin{figure}[!ht]
  \includegraphics[width=9cm]{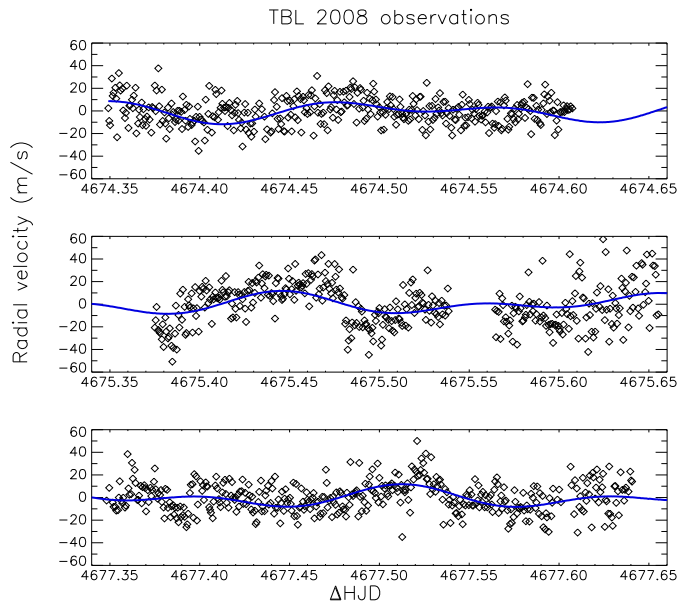}
\caption{Residual radial velocity variations of Vega (NARVAL/TBL) 2008. Superimposed
(continuous line) is the result of the corresponding frequency analysis (Table \ref{table:freqresume}).
Time is expressed in HJD = 2450000 + $\Delta$\,HJD.}
\label{fig:tstbl08}
\end{figure}

\begin{figure}[!ht]
  \includegraphics[width=9cm]{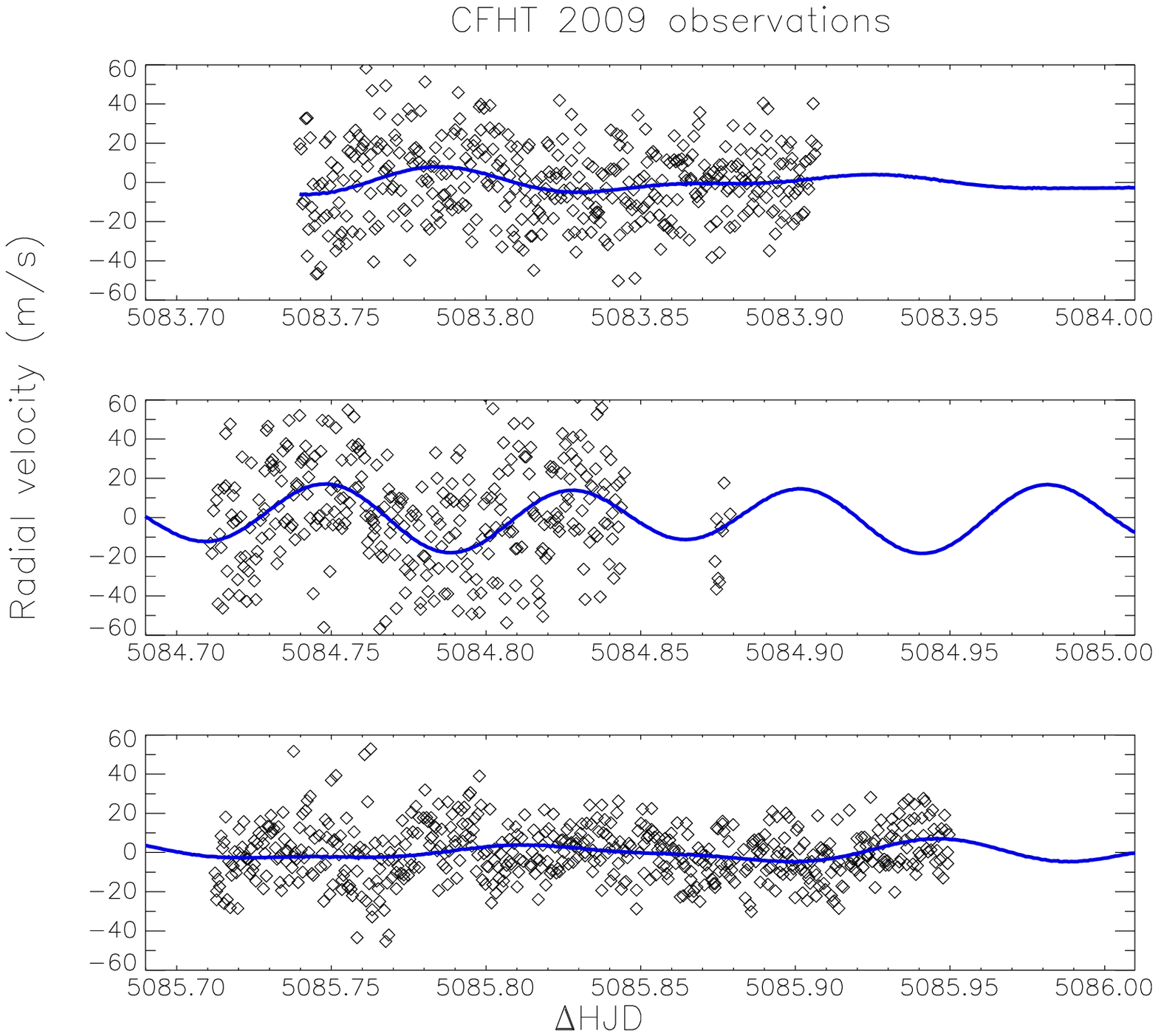}
\caption{Residual radial velocity variations of Vega (ESPaDOnS/CFHT) 2009. Superimposed
(continuous line) is the result of the corresponding frequency analysis (Table \ref{table:freqresume}).
Time is expressed in HJD = 2450000 + $\Delta$\,HJD.}
\label{fig:tscfht}
\end{figure}

\begin{figure}[!ht]
  \includegraphics[width=9cm]{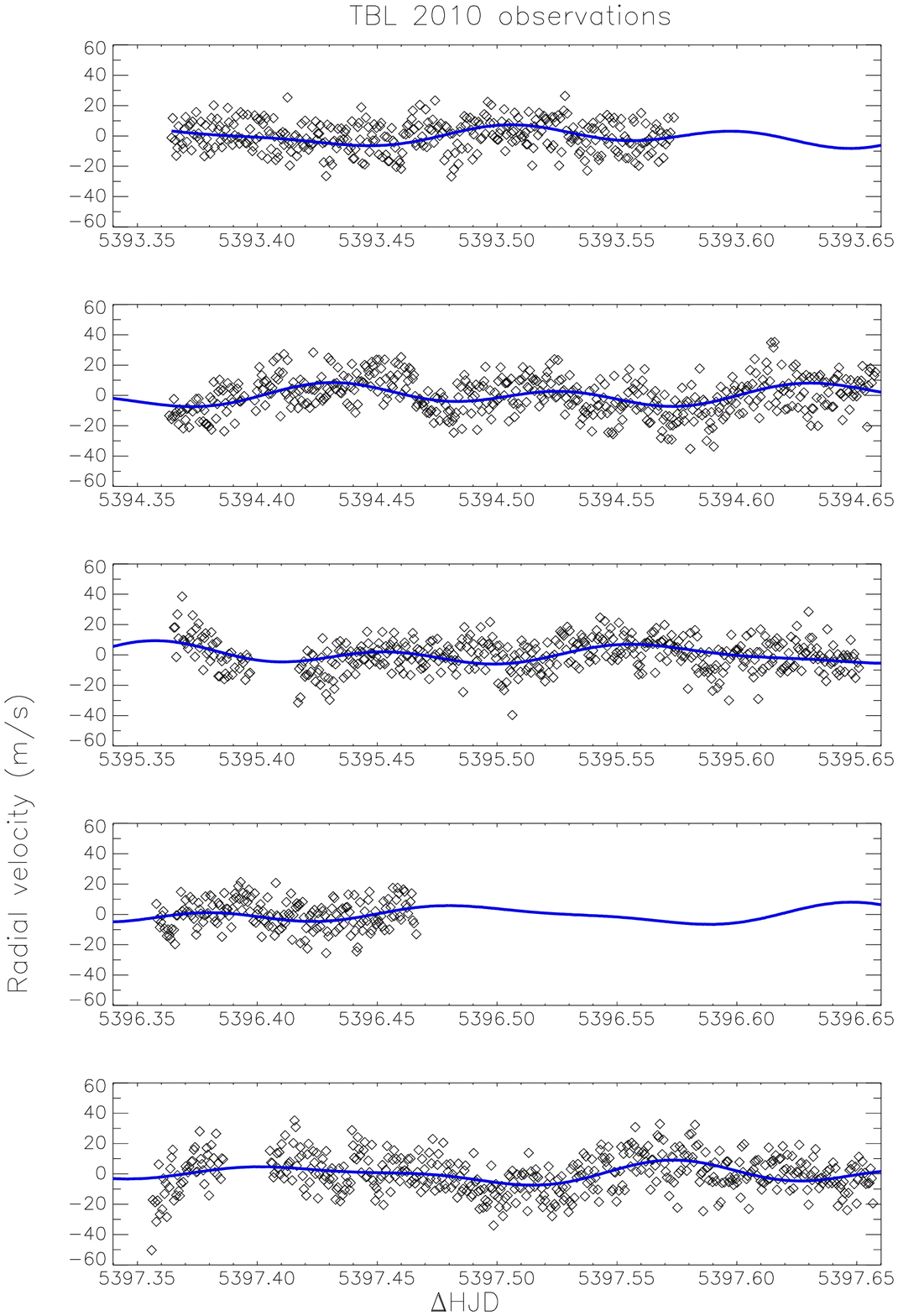}
\caption{Residual radial velocity variations of Vega (NARVAL/TBL) 2010. Superimposed
(continuous line) is the result of the corresponding frequency analysis (Table \ref{table:freqresume}).
Time is expressed in HJD = 2450000 + $\Delta$\,HJD.}
\label{fig:tstbl10}
\end{figure}

Figs. \ref{fig:tstbl08},  \ref{fig:tscfht} and \ref{fig:tstbl10} show the residual radial velocity variations of Vega observed with NARVAL/TBL (2008), ESPaDOnS/CFHT (2009) and NARVAL/TBL (2010), respectively. These curves represent the radial velocity variations after the nightly polynomial normalisation. Superimposed are the results of the corresponding frequency analysis (Table \ref{table:freqresume}) with the Period04 package \citep{lenzbreger2005}. This package is based on the discrete Fourier Transform algorithm and multi-sine least-square fits of the data. Interestingly, night 3 of the TBL (2008) data set was the only night in which data were acquired in pure spectroscopic mode; Fig. \ref{fig:tstbl08} shows well that the radial velocity curve does in no way depend on the acquisition mode, since variations are of the same order of magnitude and in phase within all three nights. This rules out the polarimetric mode as generator of faint radial velocity variations. 

\begin{figure}[!ht]
  \includegraphics[width=9cm]{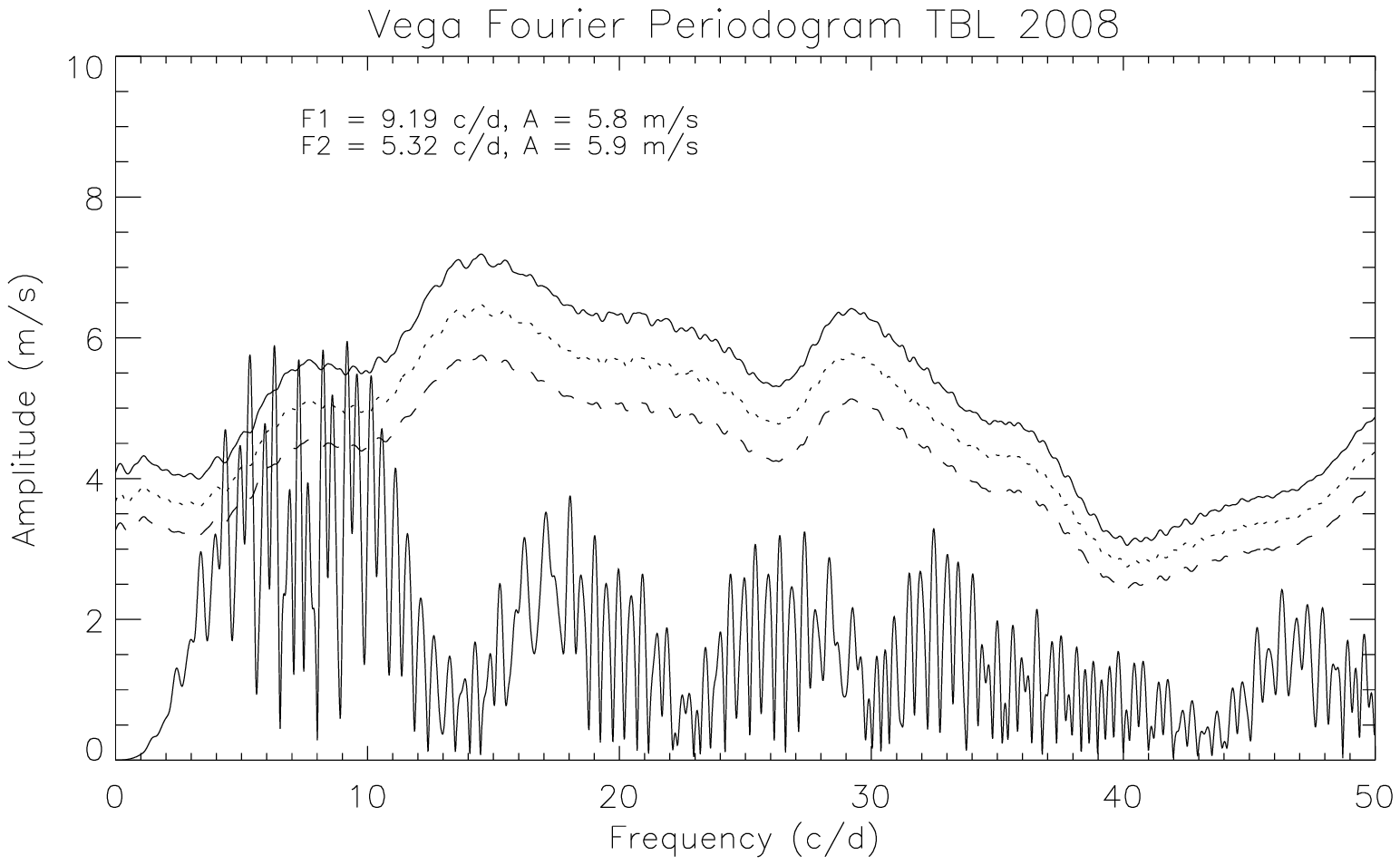}
\caption{Result of a Period04 frequency analysis on the NARVAL/TBL data of 2008. Confidence levels are: 99.9\% (cont), 99.0\% (dotted) and 90.0\% level (dashed). These levels were level calculated on a 10\,\cd\, sliding window after prewhitening with F1 and F2.}
\label{fig:F1_tbl08}
\end{figure}

\begin{figure}[!ht]
  \includegraphics[width=9cm]{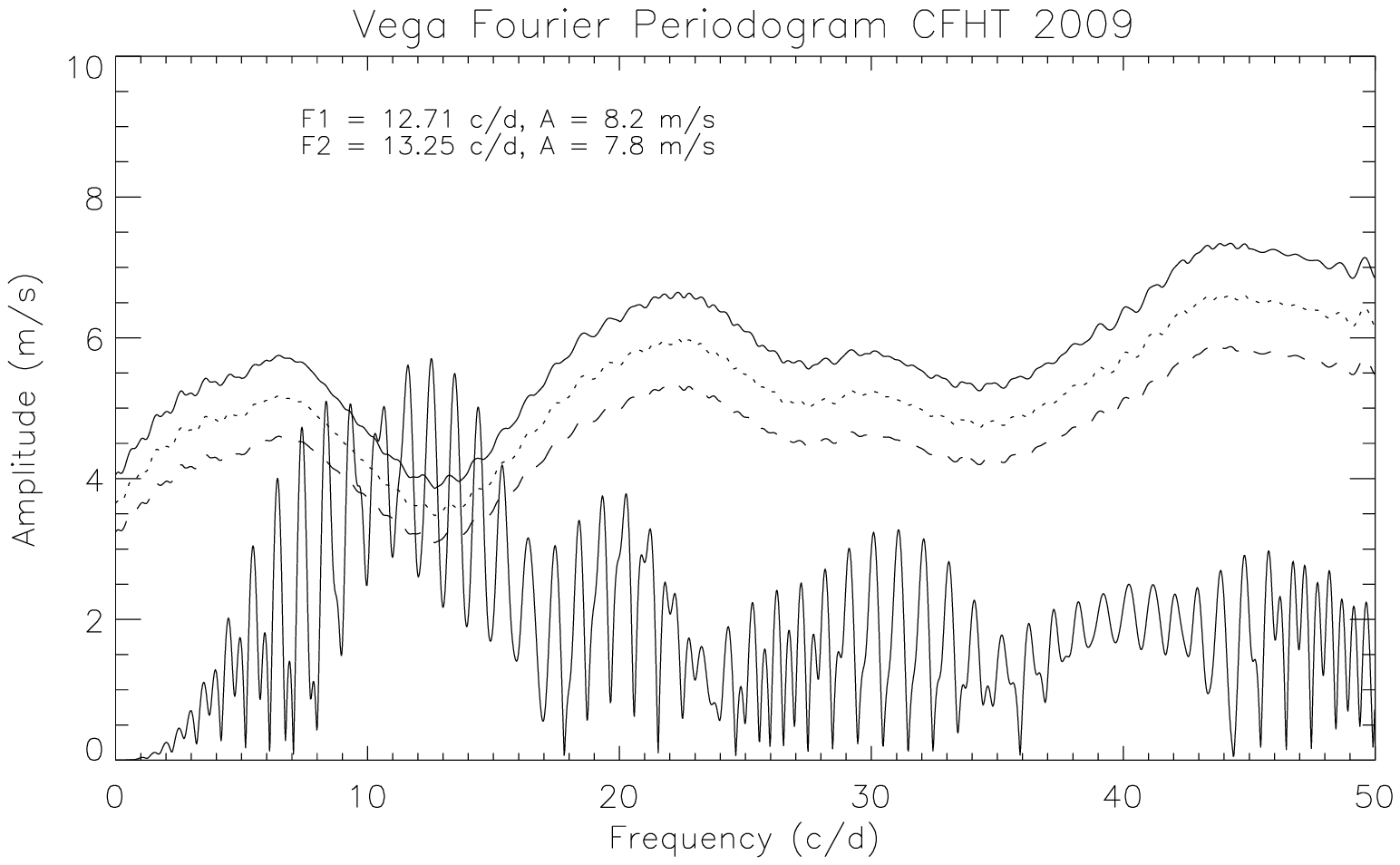}
\caption{Result of a Period04 frequency analysis on the ESPaDOnS/CFHT data of 2009. Confidence levels are: 99.9\% (cont), 99.0\% (dotted) and 90.0\% level (dashed). These levels were level calculated on a 10\,\cd\, sliding window after prewhitening with F1 and F2.
}
\label{fig:F1_cfht}
\end{figure}

\begin{figure}[!ht]
  \includegraphics[width=9cm]{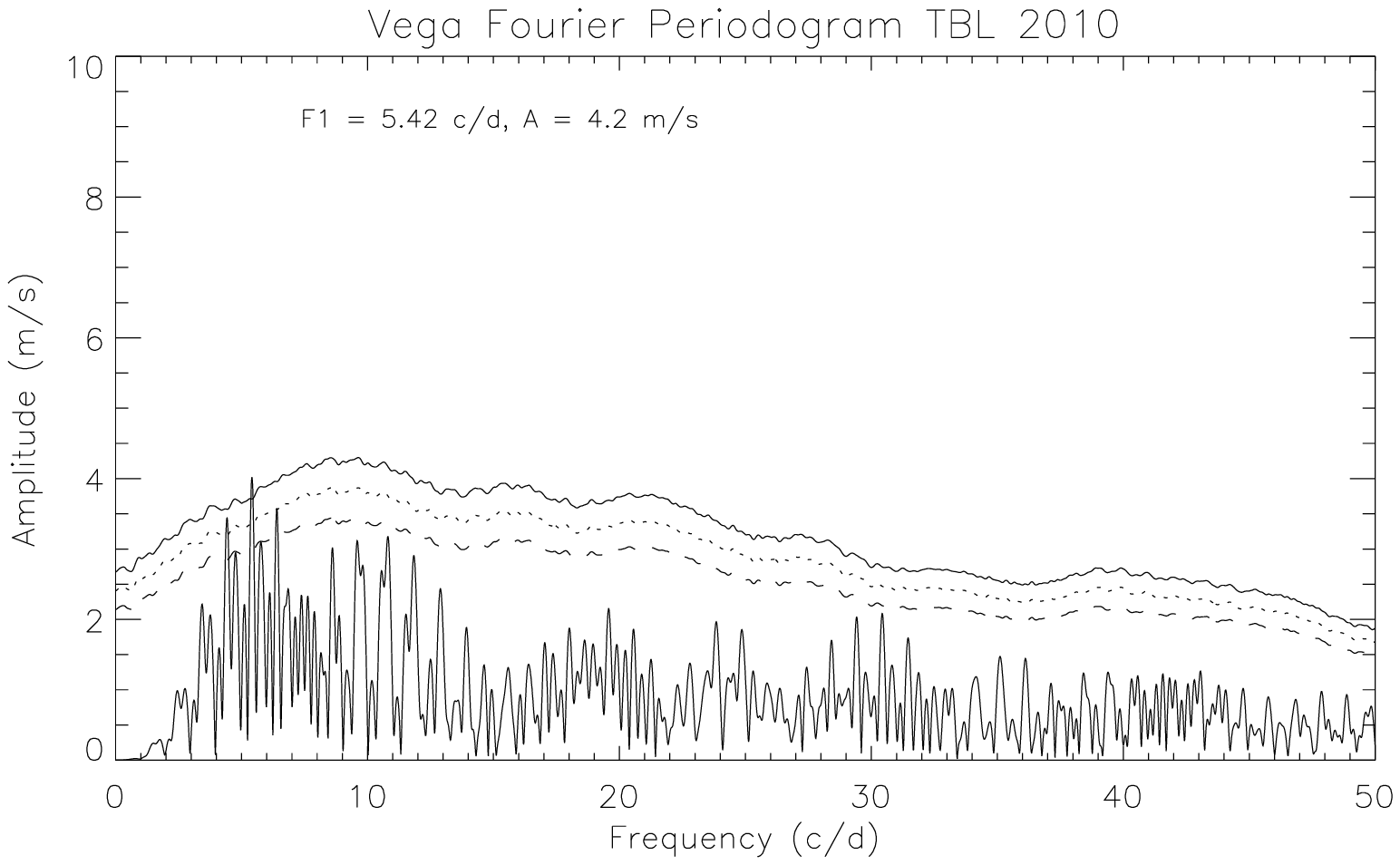}
\caption{Result of a Period04 frequency analysis on the NARVAL/TBL data of 2010. Confidence levels are: 99.9\% (cont), 99.0\% (dotted) and 90.0\% level (dashed). These levels were level calculated on a 10\,\cd\, sliding window after prewhitening with F1.} 
\label{fig:F1_tbl10}
\end{figure}

Confidence levels have been calculated according to \cite{breger1993}, who studied empirically the possibility of a peak in the periodogram being a true signal of pulsations with respect to the noise level; this work was refined later by \cite{kuschnig}: the significance of a peak in the amplitude periodogram exceeding 4.0 times the mean noise amplitude level after prewhitening of all local frequencies has a 99.9\% probability to be due to stellar pulsations  (99.0\% for a ratio of 3.6, 90.0\% for a ratio of  3.2). Figs. \ref{fig:F1_tbl08}, \ref{fig:F1_cfht} and \ref{fig:F1_tbl10} show the results of the Period04 frequency analysis for the three different runs. The two dominant frequencies in the data set of 2008 are F1$_{\rm 2008}$ = 9.19 \cd\, (A = 5.8 m\,s$^{-1}$) and F2$_{\rm 2008}$ = 5.32 \cd\, (A = 5.9 m\,s$^{-1}$). We stopped the frequency analysis after extraction of the two strongest frequencies, the next potential frequency having a significantly fainter amplitude. 
In the 2009 data set, the two dominant frequencies are F1$_{\rm 2009}$ = 12.71 \cd\, (A = 8.2 m\,s$^{-1}$) and F2$_{\rm 2009}$ = 13.25 \cd\,  (A = 7.8 m\,s$^{-1}$), the next frequency shows an amplitude of around 3.6 m\,s$^{-1}$ and was not taken into account.  Finally, in the  2010 data set variations are even fainter in amplitude with F1$_{\rm 2010}$ = 5.42 \cd\, (A = 4.2 m\,s$^{-1}$). All frequencies of the different data sets are summarised in Tab. \ref{table:freqresume}; we added a potential second very low amplitude frequency from the 2010 data set for comparison purposes.

\begin{table}[!ht]
\begin{center}
\caption{Frequencies and amplitudes of the main frequencies}
\label{table:freqresume}
\begin{tabular}{|l|c|c|} \hline
ID      & freq.    & $A $ \\
        & \cd\,      & \ms \\ \hline
F1$_{\rm 2008}$ &    9.19 & 5.8 \\
F2$_{\rm 2008}$ &    5.32 & 5.9 \\\hline
F1$_{\rm 2009}$ &  12.71 & 8.2 \\
F2$_{\rm 2009}$ &  13.25 & 7.8 \\\hline
F1$_{\rm 2010}$ &    5.42 & 4.2 \\
F2$_{\rm 2010}$ &  10.82 & 3.3 \\\hline
\end{tabular}
\end{center}
\end{table}

\subsection{Line profile variations in Vega}

\begin{figure}[!ht]
  \includegraphics[width=8.5cm]{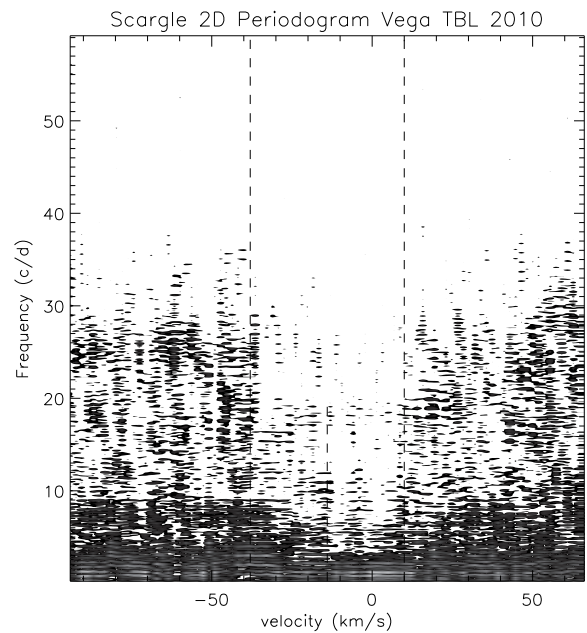}
\caption{Two-dimensional Scargle line profile analysis, corresponding to a frequency search for each velocity bin.
Data set of 2010 TBL (5 nights). The residual line profile variations have a standard deviation of 0.0007 F/Fc.
Dashed lines indicate the $\pm$\vsini limitations as well as the centroid of the photospheric equivalent profile. }
\label{fig:scargle}
\end{figure}

For almost pole-on objects like Vega, radial velocities are particularly sensitive to zonal modes of low degree $\ell$, if expressed in traditional spherical harmonics. We wanted to see if any periodic variation corresponding to a mode of higher degree $\ell$ could be seen in the equivalent photospheric profiles themselves. We therefore analysed the temporal line profile variations of the longest data set (2010) 
for each individual velocity bin of 1.8\,\kms\, by computing a 2D (velocity-frequency) Lomb-Scargle periodogram \citep{scargle,hornebal},
shown in Fig. \ref{fig:scargle}, after having subtracted the average profile of the run. It should be noted that line profile variations during this run are fainter than $\Delta F/F_{c} < 7x10^{-4}$, where $\Delta$F and F$_{c}$ are the profile variations and the (normalized) continuum flux,
respectively.  

\begin{figure}[!ht]
  \includegraphics[width=8.5cm]{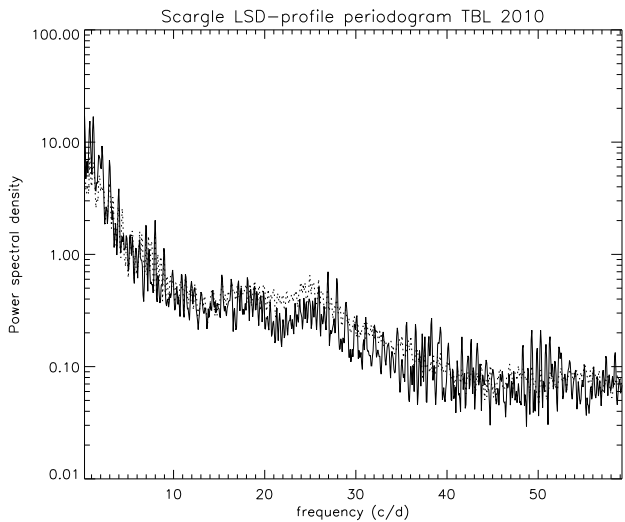}
\caption{Equivalent Scargle periodogram for the line (continous) and the continuum (dotted) during the TBL 2010 run, showing that line and continuum have the same frequency pattern, and therefore no indication of periodic line profile variations are detected at this level of precision.}
\label{fig:contraie}
\end{figure}

No indication of the presence of any particular frequency within the profile, which was not also present in the continuum, is detected. The weaker signal of the 2D periodogram within the line profile is due to the square root dependency of the noise, diminishing the absolute noise amplitude within the absorption profile.  
Fig. \ref{fig:contraie} shows the quadratic sum of the individual Scargle periodograms of all velocity bins within the  
continuum versus the sum of those within the line. Despite the logarithmic representation, no noticeable difference in the line- versus continuum-Scargle periodogram can be seen. 

\section{Discussion and conclusions}
\label{disc}

We searched for pulsations in Vega as a near pole-on configuration is particularly favorable for the seismology of rapidly rotating stars. 

Equivalent photospheric line profiles were calculated, using the multiplex information of all photospheric spectral lines in the observed wavelength 
range. In this instrumental configuration, the highest attainable radial velocity calibration was obtained by taking the telluric water vapor line profiles present in the spectrum as absolute wavelength references. A subsequent radial velocity analysis with Period04 was performed and low amplitude (A $<$ 8\ms) multi-periodic variations in the residual radial velocity curves were detected, with frequencies ranging roughly from 5 - 13\,\cd, depending on the run. The frequencies detected in this work, if confirmed, are consistent with values occasionally reported in the past, as described in the review by \citet{fernie}.

The observed amplitudes of the variations are very small with respect to the intrinsic calibration precision, but taking the telluric water vapor lines of the spectrum as a reference enabled us to reach a precision on the individual radial velocity measurement of 10-15\,ms$^{-1}$. This auto-calibration of 
the spectrum should rule out, in principle, any instrumental origin. While the spectroscopic and polarimetric set-up of NARVAL/TBL produces residual radial velocity variations of the same amplitude, and in phase with each other,  the configuration of the spectrograph itself can quite certainly be ruled out as an origin.  Nevertheless, the similarity of frequencies observed in both runs at TBL (2008 and 2010), quite different to the frequencies observed in 2009 at CFHT, leave us somewhat suspicious concerning a potential, not yet understood,  instrumental origin. 

The small amplitudes of the detected periodic variations were obtained after detrending the individual nights of each run with 2nd order polynomials. 
We must keep in mind that these nightly drifts of the radial velocity were less than 100\,\ms in the  2008 run, and less than  60\,\ms in the CFHT 2009 and TBL 2010 run, and a priori non periodic. No origin could be attributed to these variations. The positive point however was that we were able to detect the periodic signals in both, the raw and the nightly-detrended radial velocity curves. 

The intrinsic telluric line variations due to strong, high altitude wind conditions and high airmasses can be of the same order of magnitude as the detected periodicities.  These extreme values of intrinsic shifts were certainly not reached, since airmasses during our observations were always small. 
Moreover, and as an important point, the coherence of the detected periodic signals during 3 or 5 observing nights (depending on the run) and the fact that the variations are not in phase with the beginning of each night tend to rule out the telluric lines as origin of the periodic variations, and strengthen the interpretation as small amplitude stellar pulsations. 

The detected frequencies are of the same order as the fundamental mode calculated for Vega. Nevertheless, an excitation
mechanism for such modes in this region of the HR Diagram is not known yet.

At this stage we can ascertain that should Vega be confirmed as a pulsator in necessary subsequent high-precision observations with highly stable spectrographs, such as Sophie/OHP, the very small amplitudes observed during the three runs in 2008, 2009 and 2010 show that the star would belong to a category of very "quiet" pulsators. Still, as suggested by \citet{fernie}, some rarely occuring, episodic higher amplitude pulsations can not be ruled out. If pulsations were confirmed in the future, Vega would certainly be one of the most interesting stars to monitor in high-precision spectroscopy and photometry.

Another approach to search for rapidly rotating, pulsating and pole-on orientated stars is to search for such candidates within the
$\delta$-Scuti instability strip; this work has been recently initiated by our group. 

As a conclusion, one thing is certain: Vega has revealed these past years to be a more and more enigmatic star and clearly deserves our further attention!

\begin{acknowledgements}   
We thank the staffs of TBL and CFHT for their efficient help during these challenging observing runs. 
FL and TB acknowledge the support from the French ANR through the SIROCO project
GAW acknowledges Discovery Grant support from the Natural Science and Engineering Research Council of Canada (NSERC).

\end{acknowledgements}

\end{document}